\documentclass[proof]{WileyASNA-v1}

\articletype{Article Type}%

\begin{document}

\title{Oscillations of Fluid Tori around Neutron Stars}

\author[1]{Eva \v{S}r\'{a}mkov\'{a}*}
\author[1]{Monika Matuszkov\'{a}}
\author[1]{Kate\v{r}ina Klimovi\v{c}ov\'{a}}
\author[2]{Ji\v{r}i Hor\'{a}k}
\author[1,3,4]{Odele Straub}
\author[1]{Gabriela Urbancov\'{a}}
\author[1]{Martin Urbanec}
\author[2]{Vladim\'{i}r Karas}
\author[1]{Gabriel T\"{o}r\"{o}k}
\author[1]{Debora Lan\v{c}ov\'{a}}
\authormark{\v{S}r\'{a}mkov\'{a} \textsc{et al}}

\address[1]{\orgdiv{Institute of Physics}, \orgname{Silesian University in Opava}, \orgaddress{\country{Czech Republic}}}

\address[2]{\orgdiv{Astronomical Institute}, \orgname{Czech Academy of Sciences}, \orgaddress{\state{Prague}, \country{Czech Republic}}}

\address[3]{\orgdiv{ORIGINS Excellence Cluster}, \orgaddress{\state{Garching}, \country{Germany}}}

\address[4]{\orgdiv{Max-Planck-Institut f\"ur extraterrestrische Physik},  \state{Garching}, \country{Germany}}

\corres{* \email{sram\_eva@centrum.cz}}

\presentaddress{%This is sample for present address text this is sample for present address text
}

\abstract{We examine the influence of quadrupole moment of a slowly rotating neutron star (NS) on the oscillations of a fluid accretion disk (torus) orbiting a compact object the spacetime around which is described by the Hartle-Thorne geometry. Explicit formulae for non-geodesic orbital epicyclic and precession frequencies, as well as their simplified practical versions that allow for an expeditious application of the universal relations determining the NS properties, are obtained and examined. We demonstrate that the difference in the accretion disk precession frequencies for NSs of the same mass and angular momentum, but different oblateness, can reach up to tens of percent. Even higher differences can arise when NSs with the same mass and rotational frequency, but different equations of state (EoS), are considered. In particular, the Lense-Thirring precession frequency defined in the innermost parts of the accretion region can differ by more than one order of magnitude across NSs with different EoS. 
Our results have clear implications for models of the LMXBs variability.

\smallskip

%\noindent[Based on Matuszkov\'{a} et al., 2022, in preparation]
}


\jnlcitation{\cname{%
\author{ \v{S}r\'{a}mkov\'{a} E.},
\author{M. Matuszkov\'{a}},
\author{K. Klimovi\v{c}ov\'{a}},
\author{J. Hor\'{a}k},
\author{O. Straub},
\author{G. Urbancov\'{a}},
\author{M. Urbanec},
\author{V. Karas},
\author{G. T\"{o}r\"{o}k}, and
\author{D. Lan\v{c}ov\'{a}}} ,
\ctitle{Oscillations and Precession of Fluid Accretion Disks around Neutron Stars}, \cjournal{}, \cvol{}.}

%%\fundingInfo{Funding info text.}

\maketitle

%\footnotetext{\textbf{Abbreviations:} ANA, anti-nuclear antibodies; APC, antigen-presenting cells; IRF, interferon regulatory factor}

\section{Motivation}\label{sec1}

Oscillations of accretion disks have been studied in various astrophysical contexts. Among others, the analysis of disk oscillation modes has been motivated by the phenomenon of high-frequency quasi-periodic oscillations (HF QPOs) observed in the light curves of  Galactic low-mass X-ray binaries (see, e.g., \cite{kot-etal:2020} and references therein for a review). Despite the fact that a large number of these observations have been gathered over the past three decades, there is yet no commonly accepted model for either black hole (BH) or neutron star (NS) HF QPOs. Based on several strong arguments, however, it is generally believed these oscillations originate in orbital motion in the vicinity of the compact object, and numerous models have been proposed in this context, often assuming some form of accretion disk oscillations. Over the years, different scenarios have been assumed in different models incorporating different oscillatory modes and different types of accretion flow, either geodesic or non-geodesic. In this work, we focus on the possibility of epicyclic oscillations in a non-geodesic flow, which is described by the model of an equilibrium perfect fluid torus. We briefly report on some results that have been obtained within our large and detailed presently completed study.

\section{Model of an oscillating torus}\label{subsec2}

To model the torus, we assume a stationary, purely azimuthal flow of perfect fluid with constant specific angular momentum~$l=-u_\phi/u_t$. The fluid has a four-velocity $u^\mu$ with only two non-zero components,
\begin{equation}
	u^\mu = A(1,0,0,\Omega),
\end{equation}
where $A$ is the time component $u^{t}$ and $\Omega$ is the orbital velocity with respect to a distant observer. 
One may write 
\begin{eqnarray}
&A = u^{t} =  ( - g_{tt} - 2 \Omega g_{t\varphi} - \Omega^2 g_{\varphi\varphi} )^{-1/2},  \\
&\Omega = \frac{u^\varphi}{u^t} = \frac{g^{t\varphi} - l g^{\varphi\varphi}}{g^{tt} - l g^{t\varphi}}.
\end{eqnarray}
The perfect fluid with rest-mass density $\rho$, pressure $p$ and the energy density $e$ is characterized by the stress-energy tensor
\begin{equation}
T^{\mu\nu} = (p+e) u^\mu u^\nu + p g^{\mu\nu}.
\end{equation}
For a polytropic fluid, we may write
\begin{eqnarray}
p = K \rho^\frac{n+1}{n}, \,\,\,\,\,\,\,
e = np + \rho,
\end{eqnarray}
where $K$ and $n$ denote the polytropic constant and the polytropic index, respectively. In this work, we use $n=3$, which describes a radiation-pressure-dominated torus.

The Euler equation is obtained from the energy–momentum conservation law, $ \nabla_\mu T^\mu_{\; \ \; \nu} = 0 $, 
\begin{equation}
    \label{eul}
	\nabla_\mu (\ln \epsilon) = - \frac{\nabla_\mu p}{p + e}, 
\end{equation}
with  $\epsilon$  being the specific energy 
\begin{equation}
	\epsilon = - u_t = \left( - g^{tt} + 2 l g^{t\varphi} - l^2 g^{\varphi\varphi} \right)^{-1/2}.
\end{equation}
By integrating (\ref{eul}), we obtain the Bernoulli equation in the form
\begin{equation}
\label{ber}
H \epsilon = \mathrm{const}.,
\end{equation}
where $H = (p+e)/\rho$ denotes the specific enthalpy. From relation (\ref{ber}), one can easily derive equations describing the structure and the shape of the torus: 

%----------------
\begin{eqnarray}
%----------------
    \frac{p}{\rho}  &= \frac{p_0}{\rho_0} f( r, \theta ), \label{pkuro}, \,\,\,\, f( r, \theta ) = \frac{1}{nc_{\mathrm{s},0}^2} \left[ \left( 1 + nc_{\mathrm{s},0}^2 \right) \frac{\epsilon_0}{\epsilon} - 1 \right], \label{f}
%----------------    
\end{eqnarray}
%----------------
where $ c_{\mathrm{s}} $ is the sound speed in the fluid. The surface of the torus, which coincides with the surface of zero pressure, is given by the condition $f(r, \theta ) =0$.

\subsection{Epicyclic oscillations}

The concept of epicyclic oscillations within test particle motion is a well explored topic, see, e.g., \cite{Ali+Gal-etal:1981}. 
In this work, we are interested in how the properties of epicyclic oscillations, in particular the oscillation frequencies, differ for an oscillating perfect fluid torus. Such problem has been investigated in the past, first in a pseudo-Newtonian study \citep{bla-etal:2007} and later it was generalized to Kerr geometry \citep{str+sram-etal:2009}. Here we assume fluid tori around rotating NSs and extend the previous calculations for the case of Hartle-Thorne (HT) metric that well approximates the spacetime around such objects.

\section{Hartle-Thorne geometry and its parameters range relevant to rotating NSs}\label{ht}

The exterior solution of the HT metric is characterized by three parameters: the gravitational mass $M$, angular momentum $J$ and the quadrupole moment $Q$ of the star. In this work, we use the HT metric assuming dimensionless forms of the angular momentum and the quadrupole moment, $j=J/M^2$ and $q=Q/M^3$. 

The explicit form of the metric, which is too long to be presented here, may be found in \cite{urb-etal:2019} where there is also a thorough discussion of the relevance of the HT geometry for the calculations of the geodesic orbital motion. Here we just briefly summarize the appropriate ranges of the individual parameters that are implied by the present NS equations of state: The maximum value of the specific angular momentum of a NS is about  $j_{\mathrm{max}}~\sim~0.7$, the so-called oblateness parameter defined as $q/j^2$ takes values from $q/j^2~\sim~1.5$ for a very massive (compact) NS up to $q/j^2~\sim~10$ for a low-mass NS, and the conservative expectations of the NS mass values are about $1.4-2.5\,M_\odot$.

\section{Using perturbation method to calculate epicyclic frequencies in thicker tori}

As in the previous studies, we use a perturbation method to calculate the epicyclic frequencies in fluid tori. We start with the so-called Papaloizou-Pringle equation, which is a partial differential equation derived by \cite{pap+pri-etal:1984} who studied global linear stability of Newtonian fluid tori with constant specific angular momentum with respect to non-axisymmetric perturbations. The derived equation represents an eigenvalue problem that in general case has no analytic solution. It however does have analytic solution in the specific limit of an infinitely slender torus. 

This was utilized by \cite{blaes:1985} who calculated the full spectrum of modes of a Newtonian infinitely slender torus with constant specific angular momentum. The epicyclic modes are one of these modes. Later, \cite{Abr-etal:2006} derived a relativistic version of this equation, which can be written as

\begin{eqnarray} 
	&\frac{1}{\sqrt{-g}} \partial _\mu \frac{ \sqrt{-g} g^{\mu\nu} f^n \partial_\nu W }{nc_{\mathrm{s},0}^2 f +1} \nonumber + \left( l_0 \omega - m \right)^2 
	\frac{\Omega g^{t\phi} -g^{\phi \phi}}{1-\Omega l_0}
	\frac{f^{n}}{nc_{\mathrm{s},0}^2 f + 1}
	 W 	\nonumber \\ 
	& =  - \frac{2 n \aaa^2 \left( \overline{\omega} - m \overline{\Omega} \right) ^2}{\beta^2 r_0^2} f^{n-1} W, 
	\label{papa}
\end{eqnarray}
where $ \lbrace \mu,\nu \rbrace \in \lbrace r, \theta \rbrace $, $ \aaa \equiv A/A_0 $, 
$ \overline{\Omega} \equiv \Omega / \Omega_0 $,   
$ \overline{\omega} \equiv \omega / \Omega_0 $, $ g $ is the determinant of the metric tensor and $W$ equals to 
\begin{equation}
W = - \frac{\delta p}{A \rho \left( \omega - m \Omega \right)}.
\end{equation}
The subscript zero refers to quantities defined at the torus centre, $r=r_0$. 

We now expand the perturbation equation around its slender torus limit. In order to do that, we introduce a dimenssionless parameter \citep{Abr-etal:2006},

\begin{equation}
	\beta^2 = \frac{2nc_{\mathrm{S},0}^2}{r_0^2 \Omega_0^2 A_0^2},
\end{equation}

which determines the thickness of the torus. Its values range from zero to one, $\beta->0$ corresponding to the limit of an infinitely slender torus and $\beta->1$ corresponding to a torus whose outer edge extends to infinity. We expand the formulae for relevant quantities in $\beta$ by writing

\begin{eqnarray}
	&Q = Q^{(0)} + \beta Q^{(1)} + \beta^2 Q^{(2)} + \cdots, \\
	&Q \in \left\lbrace \overline{\omega}, W, \aaa, \overline{\Omega}, f \right\rbrace.
	\label{rozvoj}
\end{eqnarray}

Substituting that into the relativistic Papaloizou-Pringle equation and comparing terms of the same order in $\beta$, we obtain analytic expressions for the corrections to eigenfunctions and eigenfrequencies of the desired modes.
Doing so leads to very complicated and long expressions, to which we have found shorter approximations, but here we choose to present our results in terms of figures.

%----------------------------------------------------
\section{Results and Conclusions}
%----------------------------------------------------

\begin{figure*}[t]
\centerline{\includegraphics[width=0.8\linewidth]{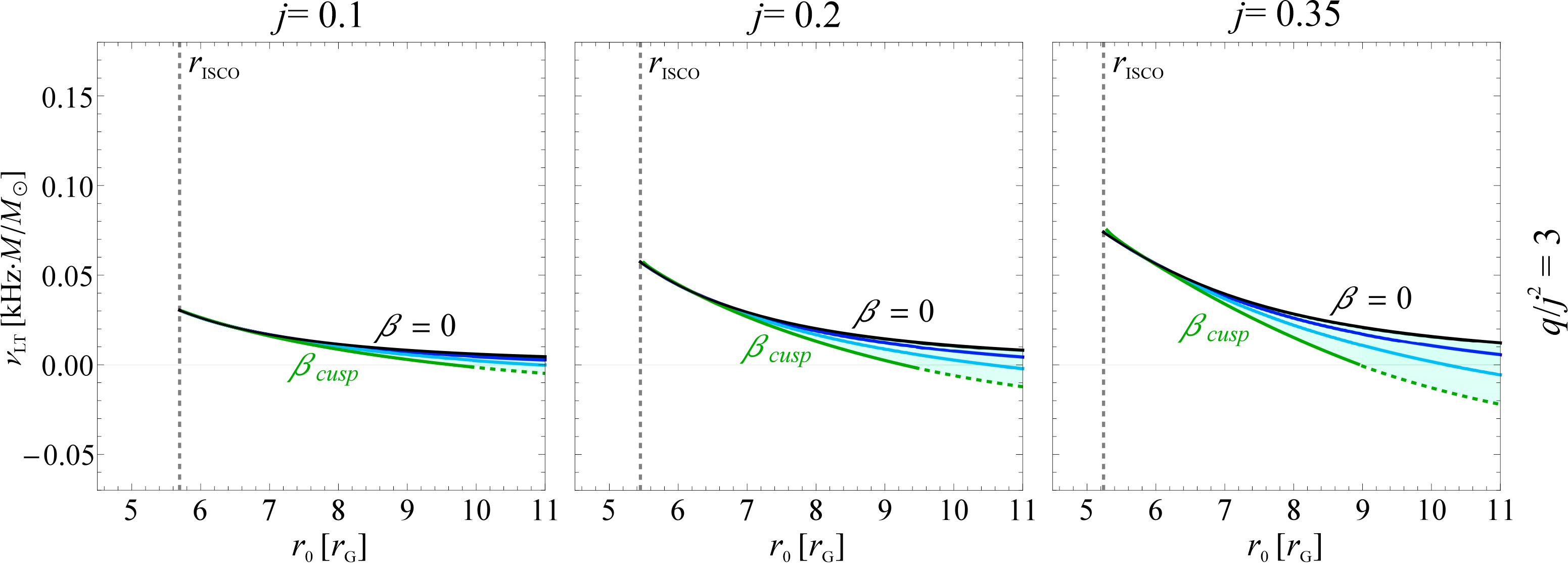}}
\caption{The Lense-Thirring precession frequency ($m=-1$ non-axisymmetric vertical epicyclic frequency) calculated for different torus thickness ($\beta$) and NS angular momentum ($j$) in the case of a small rotationally induced neutron star oblateness ($q/j^2=3$).\label{fig1}}
\end{figure*}

\begin{figure*}[t]
\centerline{\includegraphics[width=0.8\linewidth]{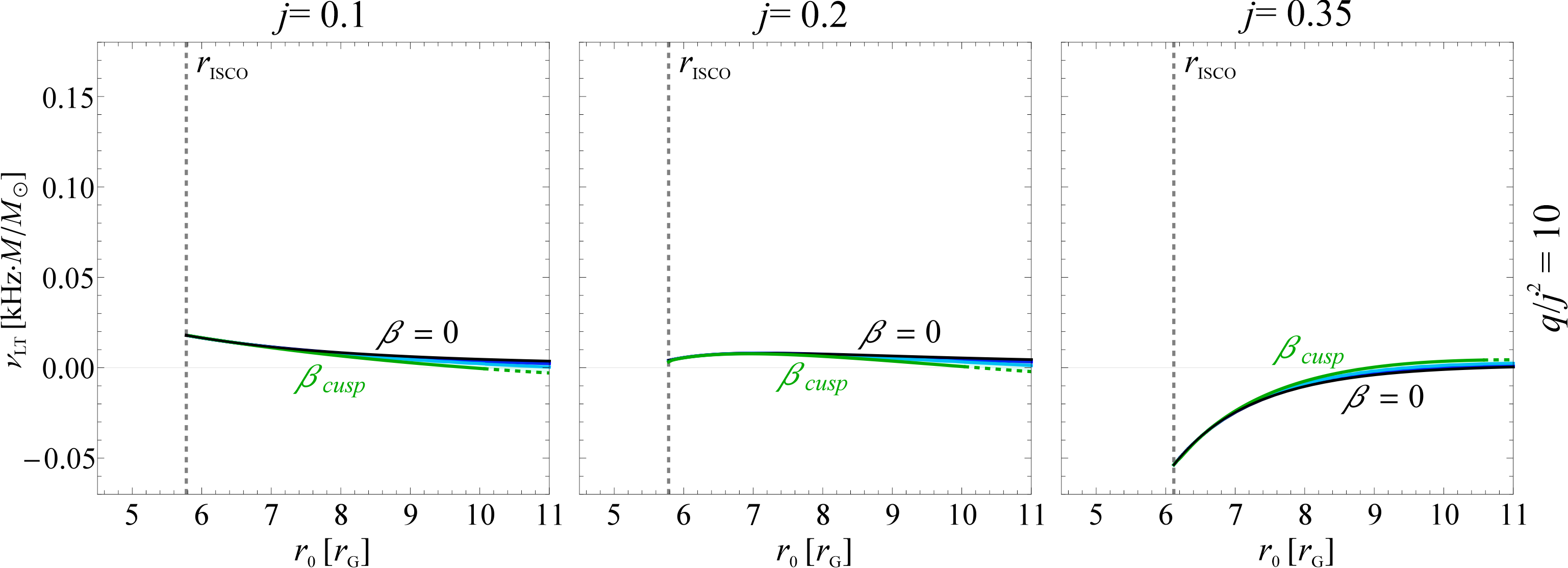}}
\caption{The same as Figure \ref{fig1} but for a large rotationally induced NS oblateness ($q/j^2=10$).\label{fig2}}
\end{figure*}

We present here the results for the $m=-1$ nonaxisymmetric vertical epicyclic mode, which corresponds to the often discussed and astrophysically relevant Lense-Thirring precession. This mode in our study possesses the most interesting features when studied from the point of view of the quadrupole moment influence. Figures \ref{fig1} and \ref{fig2} display the behaviour of its frequency calculated for different torus thickness ($\beta$) and NS angular momentum ($j$). Figure \ref{fig1} shows the frequency behaviour in the case of a small rotationally induced NS oblateness, while Figure \ref{fig2} displays the same, but for a large NS oblateness. One may see from Figure \ref{fig1} that with growing angular momentum of the star, the frequency behaviour looks qualitatively similar, but the frequencies themselves tend to increase. As we move towards higher NS oblateness (Figure \ref{fig2}), however, the effect of increased frequencies begins to fade away. This means that the presence of the quadrupole moment suppresses the effect of rotation, and for highly rotating stars we thus get similar frequency behaviour as we would for the same star with small or no rotation. Furthermore, for high NS oblateness, qualitative changes of frequency behaviour begin to arise with growing values of $j$, up to a point, where the frequencies take on negative values. This corresponds to a change of sign in the precession.

Our results indicate that, while the shape of a (non-oscillating) torus is not much sensitive to the NS quadrupole moment, the frequencies of the epicyclic modes of tori oscillations are affected significantly. In particular, the difference of frequencies of oscillations of tori around NSs of the same mass and angular momentum, but different NS oblateness, can reach up to tens of percent. Consequently, the large differences can also arise when NSs with the same mass and rotational frequency, but different EoS, are considered. For instance, the Lense-Thirring precession frequency in the innermost parts of the accretion region can differ by more than one order of magnitude among stars with different EoS.

The presented results have a clear relevance for modeling of the high-frequency quasi-periodic oscillations, in particular for models that directly deal with epicyclic modes of tori oscillations, see the work of \cite{tor-etal:2022} and references therein. The properties of epicyclic oscillations along with related astrophysical consequences will be explored in detail within our larger study that is presently being completed.

\section*{Acknowledgments}

We acknowledge the \fundingAgency{Czech Science Foundation} (GA\v{C}R) grant No. \fundingNumber{21-02685X}. We express our thanks to the \fundingAgency{INTEREXCELLENCE} project No. \fundingNumber{LTI17018}. We furthermore acknowledge two internal grants of the \fundingAgency{Silesian University in Opava}, \fundingNumber{SGS/12,13/2019}. KK thanks to the \fundingAgency{INTER-EXCELLENCE} project No. \fundingNumber{LTT17003}. MM and MU thank to the \fundingAgency{INTER-EXCELLENCE} project No. \fundingNumber{LTC18058}.  DL thanks the Student Grant Foundation of the \fundingAgency{Silesian University in Opava}, Grant No. \fundingNumber{SGF/1/2020}, which has been carried out within the EU OPSRE project entitled ``Improving the quality of the internal grant scheme of the Silesian University in Opava'', reg. number: CZ.02.2.69/0.0/0.0/19\_073/0016951.

%\nocite{*}% Show all bib entries - both cited and uncited; comment this line to view only cited bib entries;

\bibliography{Wiley-ASNA.bib}

\end{document}